\makeatletter
\newcommand{\linebreakand}{%
  \end{@IEEEauthorhalign}
  \hfill\mbox{}\par
  \mbox{}\hfill\begin{@IEEEauthorhalign}
}
\makeatother
\documentclass[conference, a4paper]{IEEEtran}
\IEEEoverridecommandlockouts
\usepackage{cite}
\usepackage{amsmath,amssymb,amsfonts}
\usepackage{algorithmic}
\usepackage{graphicx}
\usepackage{textcomp}
\usepackage{xcolor}
\usepackage{booktabs} 
\usepackage{geometry}
 \def\BibTeX{{\rm B\kern-.05em{\sc i\kern-.025em b}\kern-.08em
    T\kern-.1667em\lower.7ex\hbox{E}\kern-.125emX}}
\begin{document}

\title{Significant Ties Graph Neural Networks for  Continuous-Time Temporal Networks Modeling
}

\author{\IEEEauthorblockN{1\textsuperscript{st} Jiayun Wu}
\IEEEauthorblockA{\textit{College of Computer and Information Science} \\
\textit{Southwest University}\\
Chongqing, China \\
wujiayun@email.swu.edu.cn}
\and
\IEEEauthorblockN{2\textsuperscript{nd} Tao Jia}
\IEEEauthorblockA{\textit{College of Computer and Information Science} \\
\textit{Southwest University}\\
Chongqing, China \\
tjia@swu.edu.cn}
\linebreakand
\IEEEauthorblockN{3\textsuperscript{rd} Yansong Wang}
\IEEEauthorblockA{\textit{College of Computer and Information Science} \\
\textit{Southwest University}\\
Chongqing, China \\
yansong0682@email.swu.edu.cn}
\and
\IEEEauthorblockN{4\textsuperscript{th} Li Tao}
\IEEEauthorblockA{\textit{College of Computer and Information Science} \\
\textit{Southwest University}\\
Chongqing, China \\
tli@swu.edu.cn, corresponding author
}
}

\maketitle

\begin{abstract}
Temporal networks are suitable for modeling complex evolving systems. It has a wide range of applications, such as social network analysis, recommender systems, and epidemiology. Recently, modeling such dynamic systems has drawn great attention in many domains. However, most existing approaches resort to taking discrete snapshots of the temporal networks and modeling all events with equal importance. This paper proposes Significant Ties Graph Neural Networks (STGNN), a novel framework that captures and describes significant ties. To better model the diversity of interactions, STGNN introduces a novel aggregation mechanism to organize the most significant historical neighbors' information and adaptively obtain the significance of node pairs. Experimental results on four real networks demonstrate the effectiveness of the proposed framework. 
\end{abstract}


\section{Introduction}
The network is an effective tool for modeling social relationships, biological effects \cite{michalski2022temporal}, information systems and other complex systems. Nodes represent entities and links for the relationship between them \cite{barabasi2012network}. There is nothing is permanent except for change. The emergence and disappearance of nodes and edges are constantly acted in real networks. Compared with static networks containing fixed nodes and edges, temporal networks \cite{holme2012temporal} add the dimension of time and express the evolution of topology structure more realistically. Graph representation learning is the cutting-edge method for exploring the evolution model of these complex systems, which embed nodes by modeling the relations among entities.

\par Most graph representation methods work on static graphs with structures frozen in time \cite{perozzi2014deepwalk,grover2016node2vec,hamilton2017inductive,velivckovic2017graph}. Although such methods have achieved great success, they hardly capture or model the dynamic features and mechanisms in real dynamic networks, such as burst and memory effect \cite{goh2008burstiness}, triadic closure \cite{opsahl2013triadic}, self-exciting effect \cite{hawkes1974cluster} and so on. Recently, network representation learning on dynamic graphs has become a research hotspot \cite{skarding2021foundations}. There are two main approaches to representing and modeling dynamic networks. One way is discretizing the time that is, representing the dynamic network by \textit{discrete snapshots}. The other way is to keep all interaction records modeling it as the \textit{continuous}-time network. The methods based on discrete snapshots use multi-layer subgraphs to reflect network dynamics and each snapshot is a static graph aggregated by giving a user-specified time window \cite{chen2018exploiting,sankar2020dysat,jiao2021temporal}. However, it does not reflect the nonlinear characteristics of the network and there may be not any obvious or even any "correct" size of a time window \cite{ribeiro2013quantifying,ahmad2021tie}. To remedy these defects, Nguyen \cite{nguyen2018continuous} proposed a continuous-time network embedding method (CTDNE) which proposes a real temporal constraint of sampling orders for random walks, but it can’t model irregular time intervals which convey important information for analyzing dynamic interaction behaviors. Qu \cite{qu2020continuous} proposed a graph neural network framework (TDGNN) to aggregate and exploit the time interval of events to capture the different influences of events in different periods. Wen \cite{wen2022trend} used a GNNs framework (TREND) to model the self-exciting effect and the activity of nodes between pairs and nodes themselves. They treat each event in the network in isolation and equivalently, however, different events imply different meanings and uniqueness. For example, couples who are in a romantic relationship tend to bond closely, and temporal colleagues rarely interact after finishing a work project, i.e., different social ties have different significance.
\begin{figure}[ht]
    \centering
    \includegraphics[width=\linewidth]{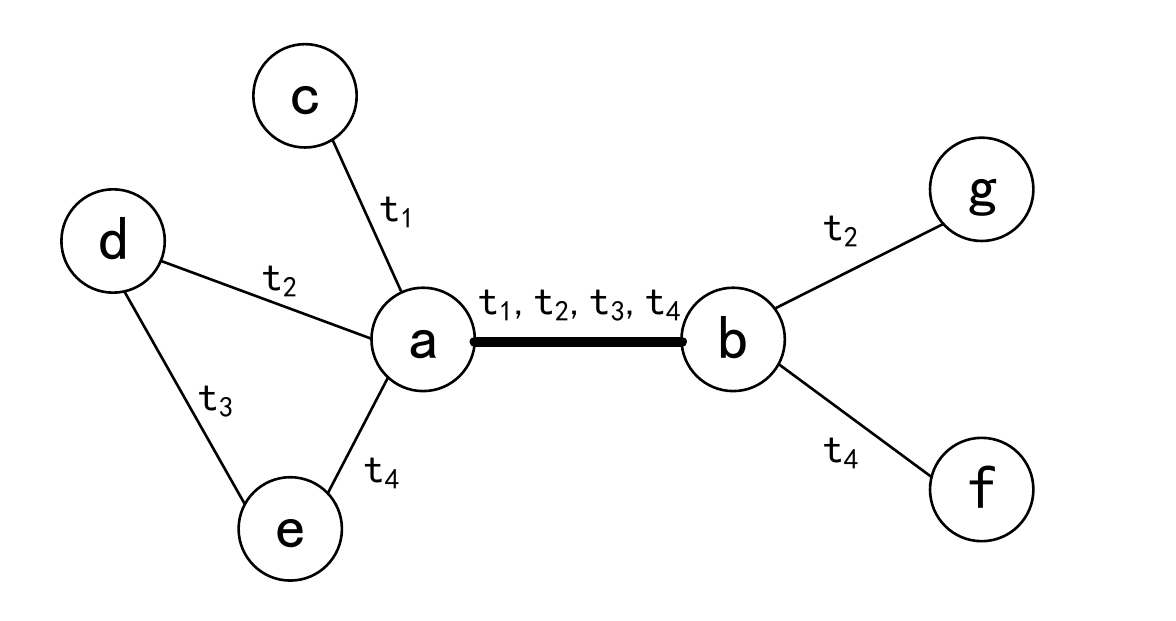}
    \caption{An illustration of the temporal network. Event (i,j,t) is the elementary unit. In this scenario, the relation between node a and b is a significant tie.}
    \label{fig:tony}
\end{figure}

\par In the temporal network \cite{holme2012temporal}, the streaming interaction events imply the evolution rules of networks. As illustrated in Fig. \ref{fig:tony}, the toy network evolving from time $t_1$ through $t_4$ can be described by a stream of triple $\{(a, b, t_1), (a,d, t_2), (a, b, t_2), (b, g, t_2), (d,e,t_3), . . .\}$, where each triple $(i, j, t)$ denotes a link formed between nodes i and j at time t. Hence, modeling and prediction of future links depend heavily on the dynamics captured in the historical link formation of each node pair. In this paper, we focus on learning the node dynamic information extraction and relation evolution, taking each event in different importance.
\par \textbf{Challenges}. To effectively model the events of link formation or network evolution on a temporal network, GNN-based frameworks have gained a lot of attention in recent years for their excellent inductive ability. However, there are still two challenges to the existing approach.

\par 1) \textit{How to capture the uniqueness of a relationship between entities?} Most GNN-based frameworks model events of the same importance \cite{qu2020continuous,wen2022trend}. In general terms, they only model whether an event $(i,j,t)$ will occur at time $t$ according to the historical information of the node pair. However, the meaning of the ties should not be deemed to the discrete and non-differential events. Take Fig. \ref{fig:tony}. as an example, three events $(a,d, t_2), (a, b, t_2), (b, g, t_2)$ happened at the same time $t_2$, however, relation between $(a,b)$ is more significant than others since node pair $(a,b)$ will continued intensive interaction after $t2$. Therefore, modeling all events in equivalent may lose the uniqueness of the relationship between entities. To more precisely describe the diversity of interactions,  a model should capture meaningful differences between future ties/relations rather than consider them as isolated and identical events.

\par 2) \textit{How to extract dynamic information as much as possible under the limits of history capacity?} Although GNN-based methods usually have excellent performance and generalization properties, the memory overhead is also high. With a limited number of aggregated units,  we can not aggregate all historical information of nodes. In real networks, some nodes have many links and interactions, i.e., the big-degree nodes. It is impractical to aggregate all the information of such nodes under a limited aggregate size. The commonly used method is to set a specific size of history (history capacity) for sampling and aggregate the history information such as \cite{chang2020continuous,qu2020continuous,wen2022trend} etc. A bigger history capacity usually leads to better accuracy, but the memory overhead grows exponentially. Therefore, how to extract more representative historical interaction dynamic features with a limited historical size for aggregation has become a bottleneck restricting performance boost.

\par \textbf{Contributions}. In this paper, inspired by the significant/strong tie effect \cite{brown1987social,lee2009online,bouchillon2014social}, we propose our GNN-based method STGNN for modeling continuous-time temporal networks. To address the problems mentioned before, To conquer challenge 1, we extract and model the significant diversity between node pairs by observing the tendency of events through the \textit{intimate window}. To optimize the mission in challenge 2, we design a significant ties aggregator \textit{(STAgg)} to extract more density information under a limited historical capacity.
Our contributions are summarized below:
\begin{enumerate}
    \item For the first time, in temporal network representation learning, we recognize the importance of modeling the diversity of ties between entities and formulate it by the \textit{intimate window}.
    \item We propose the STGNN, a novel framework with ties significance capture and historical dynamics extracting to more precisely model the relationships between ties of entities.
    \item We conduct extensive experiments on four real-world datasets, which demonstrate the advantages and availability of STGNN.
\end{enumerate}

\section{Relate Work}
In recent years,  many dynamic network representation learning models have been proposed with modeling the evolving complex dynamic systems. Typical models can be divided into three categories--temporal network embedding \cite{zuo2018embedding,du2018dynamic,lv2022graph}, graph neural network \cite{ma2020streaming, qu2020continuous, xu2020inductive, pareja2020evolvegcn,wen2022trend}, recurrent neural network based methods \cite{goyal2018dyngem, chen2019lstm, sankar2020dysat,jiao2021temporal} and etc. 
\par A dynamic network can be simplify modeled as discrete snapshots, i.e., a series of snapshots. In modeling such snapshots-based dynamic networks, the common idea is to learn node representations for each graph snapshot and then capture both the graph structures in each snapshot and the temporal effect across the snapshots. The specific key technique varies in different downstream tasks and scenarios. To be specific, mainly including matrix factorization \cite{chen2018exploiting}, skip-grams model \cite{du2018dynamic}, and RNNs. The effective and most popular tool for capturing the sequential correlation across the snapshots is recurrent neural networks (RNNs), which capture the chronological sequence of topologic variation across all snapshots. EvolveGCN \cite{pareja2020evolvegcn} evolves to upgrade GCN parameters by RNNs. CasSeqGCN \cite{wang2022casseqgcn} combines network structure and temporal sequence to predict information cascades. DynGEM \cite{goyal2018dyngem} incrementally builds the representations of a snapshot from those of the previous snapshot. DynamicTriad \cite{zhou2018dynamic} model the evolution by triadic closure process. DySAT \cite{sankar2020dysat} joint self-attention along both structural and temporal dynamics.
\par However, there are unavoidable fundamental problems in discrete snapshot models. It's hard to define a suitable size of a time window \cite{ribeiro2013quantifying,ahmad2021tie}. And since the interaction burst and memory \cite{goh2008burstiness}, it's also challenging to determine an appropriate dividing point for snapshots\cite{kivela2015estimating}. Some works explore modeling the continuous-time networks to remedy these defects of discrete snapshots. CTDNE \cite{du2018dynamic} imposed the temporal sequence constraint on a random walk to improve the accuracy. But, it ignored the connotative factor of variation. JODIE \cite{kumar2019predicting} focused on modeling the sequential interaction patterns of users and items. TGAT \cite{xu2020inductive} recognized the node embeddings with temporal features and used them for transductive and inductive tasks. TREND \cite{wen2022trend} captures events' individual and collective characteristics by integrating event and node popularity to model the dynamic process.
\begin{figure*}[ht]
    \centering
    \includegraphics[width=\linewidth]{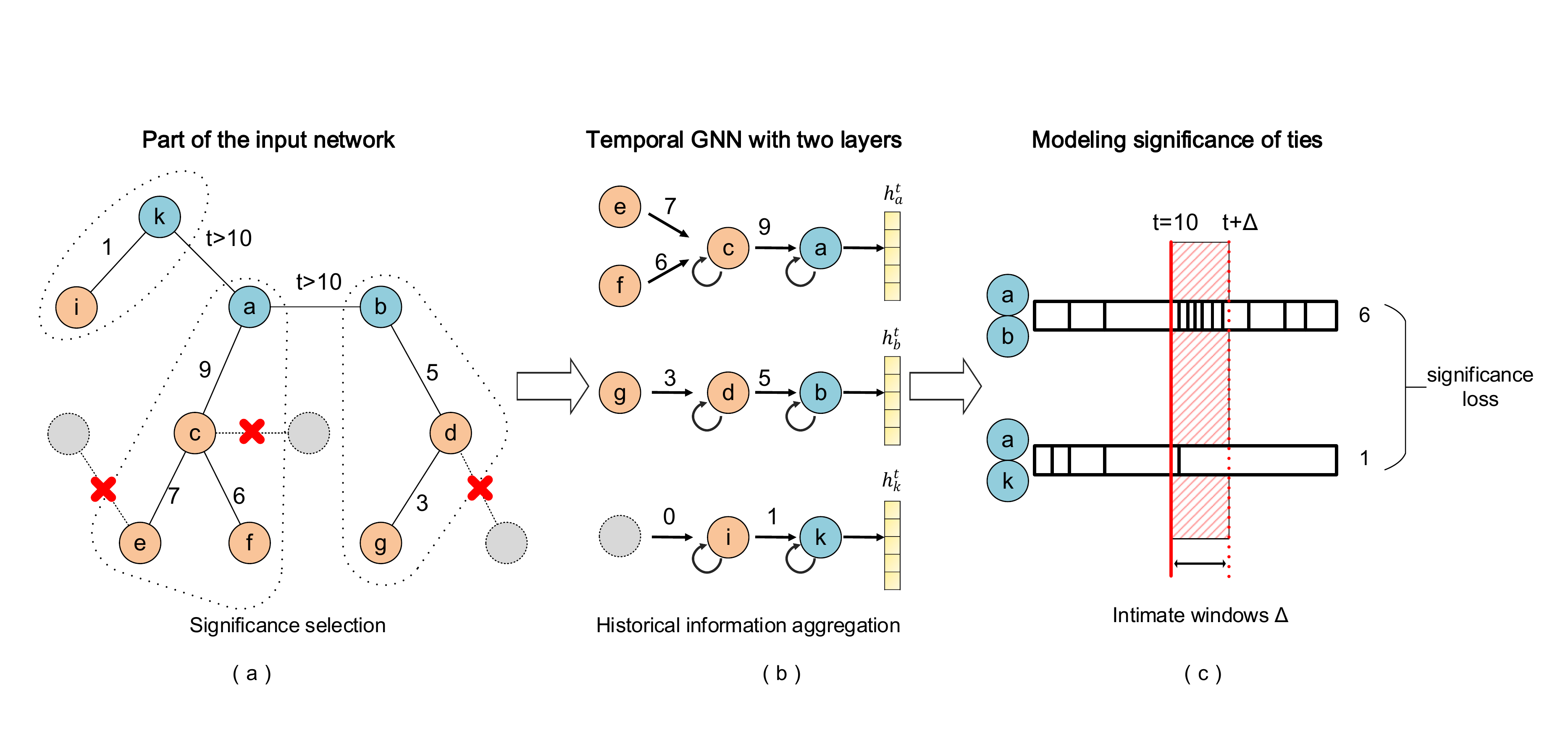}
    \caption{Overview of the STGNN. (a) the schematic diagram of significant relationship selection. (b) Two-layer temporal GNN preferentially aggregates significant neighbors. (c) Observing and modeling the significance of node pairs through the intimate window.}
    \label{fig:flow}
\end{figure*}
\section{Preliminaries}
\subsection{Problem Definition}
Given a temporal network/graph $G=\left(V, E_{T}, T\right)$, let $V$ be the set of nodes and $T$ be timestamps set, and $e^{t}_{i,j}\in E_{T}$ be the events between nodes in $V$, where $e^{t}_{i,j}$ represents node $v_{i}$ and node $v_j$ generated a contact at time $t$. 

\par In this paper, we study the temporal network representation
learning problem. Specifically, given $G=\left(V, E_{T}, T\right)$, we aim to embed network vertices into low dimensional vector space by preserving network topology dynamics, vertex content, and other edge information. It can be regard as a mapping function $g:V\to \mathcal{R}^{d}$ with $d<<N$.
\subsection{Significant ties}
Significant ties are close and strong relationships between entities. It has been extensively studied in the field of sociology. Brown \cite{brown1987social} finds that significant ties are considered more influential than weak ties and are more likely to serve as channels for information transmission.  Lee \cite{lee2009online} points out that strong social relationships lead to strong cohesion of future friendships online. Bouchillon \cite{bouchillon2014social} points out that social solid relationships foster mutual trust and allow like-minded members to connect with each other.
\par Inspired by the cross-domain research above, we build our STGNN method by capturing and modeling significant ties.
\section{Propose approach}
\subsection{Overview of STGNN}
Building upon a temporal-based GNN, the proposed STGNN can model the dynamic events, not only the generation of links but also its significance. 
\par The overall illustration of STGNN is shown in Fig. \ref{fig:flow}. First, we use the significance selection module, see Fig. \ref{fig:flow}(a) to select the most intimate neighbors of nodes as the prime candidates for the aggregation. Secondly,
as shown in Fig. \ref{fig:flow}(b), the low dimension representation vectors of nodes at time $t$ are aggregated by multiple layers of STAgg. The vectors are the input for entities' relationship significance modeling. Each STAgg layer grading extracts the neighbors' information of the prime candidates and fuses them with the node self-information. Next, we model the significance of the relationship between entities through the intimate window module, see Fig. \ref{fig:flow}(c). The observation windows after time $t$ can provide us with the intimacy level of a node pair.
\par In conclusion, we use a more effective information extraction mechanism to aggregate higher density dynamic features in a limited history capacity and endow each event with an exact significance level to more appropriately express the diversity of relationships between entities. Finally, with the benefit of the above innovation, STGNN can model the dynamic process of link generation more precisely.
\subsection{Significant Ties Aggregator}
\textbf{Significant selection} Temporal networks display non-trivial properties, which may have a profound effect on the dynamic processes deployed on them.  The inter-event, the number of contact and activation times between a node pair has been found to follow a heavy-tail or power-law distribution in many temporal networks \cite{adamic2000power, min2011spreading,grinstein2008power,jo2012circadian}. The size of node historical interaction and the interaction times with neighbors are highly imbalanced. Only a few neighbors are intimate friends of a node, i.e., one may meet many people, but the intimate partner is the minority. The nodes' most crucial historical context characteristics are determined by these intimate nodes, i.e., the significant ties. Inspire by this assumption, we pre-select possible high significant neighbors as the top candidates for information aggregation to obtain higher-density historical information.
\par For a node pair $(u,v)$ and their historical interaction records before time $t$ $\{(u,v,t_1),(u,v,t_2),...,(u,v,t_n)\}$. The initial significance level is calculated by describing the self-exciting effect and paying more attention to nearby events. The initial significance level $s^{t}_{uv}$ of the node pair $(u,v)$ at time $t$ is given by using the decay coefficient term in the standard exponential form.
\begin{equation}
\label{eq:possion}
    \begin{aligned}
    s^{t}_{uv}=\sum \limits_{i=1}^{n}e^{- \lambda(t-t_{i})},\ \lambda=1.
\end{aligned}
\end{equation}
Where $n$ is the number of interactions between node pair $(u,v)$ before time $t$, the historical influence's decay is characterized by decay coefficient $\lambda$. Note that learning a proper decay coefficient through temporal GNN is not our pursuit but modeling the significance of ties at a given time $t$, $s^{t}_{u,v}$. We give a default value of $\lambda=1$, i.e., the standard Poisson distribution.  Although the formula is simple, it already describes preliminary temporal features, i.e., expresses historical influences according to time intervals and pays more attention to recent events. It can also be regarded as a Hawkes process with an exponential function as the kernel \cite{hawkes1974cluster}. Apply this calculation to each historical neighbor of node $u$ we get the initial significance ties list $\mathbf{S}^{t}_{u}$ of node $u$ at time $t$, $\mathbf{S}^{t}_{u} = \left[s^{t}_{u1},s^{t}_{u2},...,s^{t}_{uv} \right]$. For each node, we rank their neighbors according to the initial significance, then an ordered list of prime candidates of intimate neighbors ready for the subsequent aggregation of neighbor history information.
\par \textbf{Temporal GNN layer}. For the aim of deducing the node representation at any given time, we use temporal GNNs, owing to their inductive nature. Following the message pass process, the representation of nodes at each layer is aggregated and mapped from its intimate neighbor's portraits (i.e., feature and embedding). In each layer, we do not pass every single event $(u,v,t)$ but the significance of the node pair. The representation of a node at time $t$ is the fusion of self-information and neighbor information. We aggregate the neighbor information according to their significance level. Formally, let $\mathbf{h}_u^{t, l}$ be the embedding vector of node $u$ at time $t$ with size $d_l$ in the $l$-th layer. $\mathbf{h}_u^{t, l} \in \mathbb{R}^{d_l}$ can be extracted by
\begin{equation}
\label{eq:layer}
\begin{aligned}
\mathbf{h}_u^{t, l}=\sigma\left(\mathbf{h}_u^{t, l-1} \mathbf{W}_{\text {self }}^l+
\underbrace{\sum_{ v_{i} \in \mathcal{S}^{t}_{u}(m)} \mathbf{h}_{v_{i}}^{t, l-1} \mathbf{W}_{\text {nbr }}^{l}\mathcal{\phi}(s^{t}_{uv_{i}},\beta_{i})}_{\text{significant neighbors' information}}\right)
\end{aligned}
\end{equation}
where $\sigma$ is an activation function e.g., ReLU. Both $\mathbf{W^{l}_{\text{self}}},\mathbf{W^{l}_{\text{nbr}} \in \mathbb{R}^{d_{l-1}\times d_{l}}}$ are two learnable matrices of the same size used to map the self-information and neighbor embedding information respectively in the $l$-th layer. $m$ is the capacity of history for the GNN aggregation layer. $\mathcal{S}^{t}_{u}(m)$ is an ordered list of  $m$ most significant neighbors of node $u$ at time $t$. $v_{i}$ is the $i$-th significant neighbor of node $u$ at time $t$. $\mathcal{\phi}(s^{t}_{uv_{i}},\beta_{i})$ captures the real significance of the $i$-th initial intimate neighbor with softmax. 
\begin{equation}
\begin{aligned}
\mathcal{\phi}(s^{t}_{uv_{i}},\beta_{i}) = \frac{exp\left(s^{t}_{uv_{i}} \times \beta_{i}\right)}{\sum_{v_{i} \in \mathcal{S}^{t}_{u}(m)}exp\left(s^{t}_{uv_{i}} \times \beta_{i}\right)}
\end{aligned}
\end{equation}
$\beta_{i} \in \vec{\beta}$ is a learnable correction term for the corresponding $i$-th initial significance level $s^{t}_{uv_{i}}$, the size of $\vec{\beta}$ is $m$, e.g., $\beta_{1}$ is the correction value of the most significant neighbors after inital calculation, $\beta_{m}$ for the $m$-th.
\par Overall, the aggregated information includes two parts--self-information and neighbor information. Self-information is the basic attribute of a node, which does not change over time. The aggregation of neighbors' information is indicated by the node's most intimate $m$ neighbors instead of randomly sampling from the node's historical interactions. 
In this way, we can model different significance levels ties separately and obtain information with more individual characteristics.
\par To enhance the representative capacity, we stack two temporal GNN layers. In the first layer, the node representations can be initialized by the input node features $\mathbf{X}$; in the output layer, the output representation is denoted as $\mathbf{h^{t}_{u}\in \mathbb{R}^{d}}$ denoted for node $u$ at time $t$. The collection of parameters all layers is $\theta_{g}=\left\{\mathbf{W}^{1}_{\text{self}},\mathbf{W}^{1}_{\text{nbr}},\mathbf{W}^{2}_{\text{self}},\mathbf{W}^{2}_{\text{nbr}}, \vec{\beta}\right\}$.
\subsection{Modeling Significance of Ties}
\textbf{Intimate windows}. Many temporal GNN-based methods treat events $(u, v, t)$ in isolation and equivalence \cite{qu2020continuous,wen2022trend}. They infer attribution at the event-case level. In other words, their model mainly concludes whether the node pair $(u, v)$ will be connected at time $t$ by aggregating historical information. However, they ignore the uniqueness of the relationship of entities. As shown in Fig. \ref{fig:flow}(c), both the two node pairs $(a,b),(a,k)$ will be contacted after the $t_{10}$ moment, and their meanings are equivalent if we only pay attention to whether there is an event at that moment $t_{10}$. But, if we further observe their event clusters after time $t10$, during the window period $(t_{10},t_{10}+\Delta)$, $(a, b)$ has six interactions, but $(a, k)$ has only once, which indicates that the intimacy of $(a, b)$ is more significant than $(a, k)$, i.e., the significance of relationships can be diversity. To model the significance of relationships or ties, we propose setting an intimate window with a size of $\Delta$ for observation of the future event clusters.
\par To unify the window size settings for different networks,  we quantitatively give a method for setting the size. We fit the time interval of node pairs of each network to a power-law distribution i.e., $y=x^{-\alpha}$. It perceives the time interval characteristics of interaction patterns on the temporal networks (using the Python toolkit \textit{powerlaw} \cite{alstott2014powerlaw}). According to the fitted complementary cumulative distribution function with parameters $\alpha,xmin$, $xmin$ is the lower limit of the fit range, $\alpha$ is the power. We establish a mapping function $\delta(p)$ to obtain the expected size for observing the $p$ proportion of the total events.
\begin{equation}
    \delta(p) = \left(\frac{\alpha-1}{c(1-p)}\right)^{\frac{1}{(\alpha-1)}}
\end{equation}
where $c$ is a constant related to the minimum fit value $x_{min}$, $c=(\alpha-1)x_{min}^{1-\alpha}$. $p$ is the proportion of the number of events expected to be covered by the window size $\Delta$.  The total number of interactions between node pairs during the 
intimate window $[t,t+\Delta)$ is their significance level at time $t$ represented by $s^{\Delta}_{(u,v,t)}$.
\par \textbf{Significance loss}. Combined with the intimate window module, we design a significance loss function to capture the relationship significance of node pairs. Our loss function is based on the Cosine embedding loss function, and the loss values of positive and negative samples will be calculated separately.
\begin{equation}
L_{e^{(u,v,t)}}= 
\begin{cases}
\left(1-\cos \left(\mathbf{h}_u^t, \mathbf{h}_v^t\right)\right)\times s^{\Delta}_{(u,v,t)}, & \text{if}\ s^{\Delta}_{(u,v,t)}\ge1 
\\ \\
\max \left(0, \cos \left(\mathbf{h}_u^t, \mathbf{h}_v^t\right)\right)\times \overline{s}^{\Delta}, & \text{if}\ s^{\Delta}_{(u,v,t)}=0
\end{cases}
\end{equation}
where the balance factor $\overline{s}^{\Delta}_{t}$ of is the mean value of the $s^{\Delta}_{(u,v,t)}$ among positive samples. For the loss of positive samples, the significance of the node pair $s^{\Delta}_{(u,v,t)}$ is integrated into the formula, and the node pair with a significant relationship will be given more weight. If we only apply this formula, the loss will be more inclined to positive samples, which may lead to overfitting.  To balance the weight of positive and negative samples, in the negative sample loss, we correct the loss of negative samples as the product of it and the mean significance $\overline{s}^{\Delta}$. 
\section{Experiments}
\subsection{Network data}
Four public real networks are selected to cover the diversity of networks, as described in Tab. \ref{tab:network statistics}.
\par 
 Radoslaw-email(\textbf{Radoslaw}), is another email network of internal email communication network among employees of the Radoslaw manufacturing company. Ia-contact(\textbf{Ia-Contact}), an extensively studied publicly available datasets which recorded a face-to-face human contact, but the relevant detail description is missing. College-Messages(\textbf{College}), the network includes the contact of the 1899 students in the University of California through Facebook. Wiki-election(\textbf{Wiki}), the network is based on all admin election and voting histories extracted from the edit history of Wikipedia pages.
\begin{table}[ht]
  \caption{Basic statistics of temporal networks.}
  \label{tab:network statistics}
  \resizebox{\linewidth}{!}{
  \begin{tabular}{cccccc}
    \toprule
    \textbf{Network} & $N$ & $E$ & $C$ & $D$& $T$ \\
    \midrule
    Radoslaw& 167   & 3215  & 82.9k & 25.5 & 271.2 \\
    Contact & 273   & 2123  & 28.2k & 13.3 & 3.1 \\
    College & 1899  & 13.8k & 59.8k & 4.3 & 193.7 \\
    Wiki    & 7118  & 100.8k& 107.1k& 1.06 & 1378.3 \\
  \bottomrule
\end{tabular}
}
\end{table}
\par $N$ is the number of nodes, $E$ is the number of unique edges, $C$ is the number of contacts, $D$ is the average amount of the events occur on an edge, $T$ is the total observation time count by day.
\par The basic statistics of the above network are listed in Table \ref{tab:network statistics}.
The above data sets can be obtained through the \emph{NetworkRepository} \cite{nr}.
\subsection{Link prediction task}
Link prediction is one of the most popular downstream task of network representation learning, which can reflect the model's ability of capturing dynamic features \cite{skarding2021foundations,yijun2022maximum}.
\par \textit{Temporal link prediction (TLP):} Given a temporal network $G=\left(V, E_{T}, T\right)$ and the last observation time $t_0$, $t_0 \in T$. TLP aims to predict whether a node pair $(u,v)$ will generate a link or not when $t>t_0$ based on the history information, i.e., all events $e^{t}_{i,j}\in E_{T}$ before observed time $t_0$ can be use for prediction aim. 
\subsection{Baselines}
We selected five recent novel methods as benchmarks to explore the predictive performance of our STGNN models. For static network method, we compared Node2vec with GraphSAGE. Three dynamic methods, CTDNE, SGNN, and TREND all work based on continuous-time temporal networks. The baselines are introduced as follows:
\begin{itemize}
\item \textbf{Node2Vec}\cite{grover2016node2vec}: Similar to Word2vec, a biased random walk is performed in the network to generate a walk sequence, and the skip-gram model is used to obtain the node embedding.
\item \textbf{GraphSAGE}\cite{hamilton2017inductive}: It includes two modules of sampling and aggregation. First, the topology information between nodes is used to sample the neighbors, and then the information of adjacent nodes is continuously fused together through a multi-layer aggregation function.
\item \textbf{CTDNE}\cite{nguyen2018continuous}: A method based on the temporal random walk. The temporal random walk sequence which conforms to temporal logic and is put to the skip-gram model, and the embeddings containing both structural and temporal information are obtained.
\item \textbf{SGNN}\cite{ma2020streaming}: The method regards temporal networks as streaming data and uses a message passing mechanism for representation learning which updates node information by capturing the sequential information of edges (interactions), the time intervals between edges, and information propagation coherently.
\item \textbf{TREND} \cite{wen2022trend}: A GCN-based method inspired by the self-exciting effect of the Hawkes process. It captures the individual and collective characteristics of events by integrating both event and node popularity, driving more precise modeling of the dynamic process.
\end{itemize}
\subsection{Experiments settings}
\par We divided the training and testing windows with a ratio of 3:1. Let the total time span of a network be $T$, all interactions of the period $[0, \frac{3}{4}T]$ constitute the training set and we aggregated the remaining interactions of period $(\frac{3}{4}T, T]$ to a static network as our test set. In other words, all interaction history before $ \frac{3}{4}T$ can be used to predict whether a node pair $(u,v)$ will generate a link after $ \frac{3}{4}T$.
\par For all these baselines, we set the output embedding dimension $D$ of all baselines to $D=128$. The mean average precision (\textbf{MAP}) and area under the ROC curve (\textbf{AUC}) are our evaluation metrics. The ratio of positive and negative examples is 1:1.
\par Specifically, for the static method (Node2Vec,GraphSAGE), considering the weak tie effect \cite{lu2010link}, we aggregated the training interactions into an unweighted graph as data input. We set both the sampling hyperparameters $p$ and $q$ of Node2Vec to 1. For GraphSage, the maximum number of 1/2/3/4/5-hop neighbor nodes was set to be 25/10/10/10/10. Temporal neighbor selection for CTDNE was set to unbiased distribution as suggested in \cite{nguyen2018continuous}. As suggested by the authors, we set the maximum propagation size to 50 for SGNN. TREND can output the network embedding results at any given time, and we use the embedding results at the last moment $T$ as a comparison. 
\par \textit{Settings of STGNN}. We stack two layers with a ReLU activation STAgg based temporal GNN layers. Both the dimension of hidden layer and output layer are set to 16 on all datasets. The initial features of are randomly generated vectors with a size of 128 with a value between $[-1,1]$. Two hyperparameters $m$, $p$, the history capacity $m$ is set to be $m=10$, and for all datasets. $p=0.8$ for College and $p=0.5$ for others.  We use the Adam optimizer with a learning rate 0.01. Positive and negative sample sampling is balanced. 
\par And for each methods, 10 replicate experiments were performed for each method, and the average performance is reported. For each node pair $(u,v)$, we use the follow three \{Cos, Had, L2\} similarity to measure and rank the score of node pairs, see Tab. \ref{tab:similarity}. For each methods, the best result among the three similarity is reported.
\begin{table}[ht]
    \centering
    \caption{The similarity evaluation for node pair score}
    \label{tab:similarity}
    \renewcommand{\arraystretch}{1.5}
    \begin{tabular}{cc}
        \toprule
        Evaluation & Definition\\
        \midrule
         Cosine (Cos)   &  $\frac{h_{u}\cdot h_{v}}{||h_{u}||||h_{v}||}$   \\
         hadamard(Had)  & $ sum(h_{u}\odot h_{v})$       \\
         Weighted-L2(L2) &   $|h_{u}- h_{v}|^{2}$        \\
        \bottomrule 
    \end{tabular}
\end{table}
\subsection{Overall performance}
\begin{table*}[ht]
\centering
\label{tab:preformance}
\caption{Performance compassion of link prediction}
\renewcommand{\arraystretch}{1.5}
\setlength{\tabcolsep}{1 mm}{
\begin{tabular}{ccccccccccccc}
\hline
 & 
\multicolumn{2}{c}{Radoslaw} & & 
\multicolumn{2}{c}{Ia-contact} & & 
\multicolumn{2}{c}{College} & & 
\multicolumn{2}{c}{Wiki} 
\\
\cline{2-3} \cline{5-6} \cline{8-9} \cline{11-12}&
\multicolumn{1}{c}{AUC} & \multicolumn{1}{c}{MAP} & & \multicolumn{1}{c}{AUC} & \multicolumn{1}{c}{MAP} & & \multicolumn{1}{c}{AUC} & \multicolumn{1}{c}{MAP} & &
\multicolumn{1}{c}{AUC} & \multicolumn{1}{c}{MAP} 
\\ 
\hline
 Node2Vec & 0.7840 & 0.7357 & & 0.8858 & 0.8483 && 0.7293 & 0.7468 && 0.6648& 0.6616\\ 
 GraphSAGE& 0.6488 & 0.5766 & & 0.8322 & 0.7322 && 0.7252 & 0.6962 && 0.6844& 0.6614\\
 CTDNE    & 0.7469 & 0.7462 & & 0.8209 & 0.8144 && 0.7209 & 0.8191\dag && 0.6635& 0.7427\dag\\
 SGNN     & 0.8141\dag & 0.7568\dag & & 0.8942 & 0.8864 && 0.6142& 0.6382 && 0.4946 & 0.4968 \\
 TREND    &0.7778 & 0.7335 & & 0.9288\dag & 0.9247\dag && 0.7678\dag & 0.7426 && 0.7096\dag & 0.6864 \\
\textbf{Our STGNN} & \textbf{0.8212} & \textbf{0.7786} & & \textbf{0.9461} & \textbf{0.9414} && \textbf{0.8600}& \textbf{0.8663}&& \textbf{0.7758}& \textbf{0.7640} \\
\hline
\end{tabular}}
\end{table*}
The AUC and MAP values are reported in Tab. \ref{tab:preformance}. In general, no single method can maintain excellence on all datasets. Our STGNN is the most commendable method on almost all datasets. It achieved the best AUC and MAP values on the above datasets.
\par First, compared to static methods. Our STGNN shows a significant performance boost among all datasets. This is mainly because the static methods work on the aggregation network, which doesn't contain any temporal information. Some edges might have a lower influence or disappear. Thus, the structure's information can be noisy or outdated. But, note that the performance of the static method is not necessarily inferior to the dynamic one. If the network's topology is changeless and stable within the observation range, the static method can be a good predictor,  e.g., In Ia-contact,  Node2Vec is better than CTDNE.
\par Second, compared with dynamic methods. STGNN also has significant performance improvements, and the three continuous-time-based methods (i.e., CTDNE, SGNN, TREND) correspond to three mainstream embedding ideas. CTDNE extracts features from temporal walk sequences. However, the temporal random walk paths do not reflect real-time intervals. Comparing CTDNE with Node2Vec, they both are skip-gram-based models. The performance improvement of CTDNE is may due to it extracting more useful temporal structural features, but the time sequence features are limited in an accuracy improvement. SGNN implements the message passing mechanism through LSTM, but it needs to set a suitable propagation scale. The generalization properties are limited, e.g., on the Wiki network, the method's performance approximates random selection. The inductive properties of GNN bring better generalization properties. TREND is a method based on the temporal Hawkes process GNN, which uses the GNN to model the conditional strength of node pairs at time $t$. Still, the Hawkes process usually requires a large amount of history to derive the conditional strength. Under the limited history capacity, sampling-based event selection loses a lot of feature density. Instead of acting directly on each dispersed event, STGNN models the significance of the relationship with higher information density and extracts more dynamic and structural features with a limited historical capacity. Besides, finer-grained modeling of the importance of nodes' future relationships allows us to capture further the pattern between historical events and future event clusters.
\subsection{Ablation study}
To explore the role of modules in STGNN. We design three forms ablated models on the task of temporal link prediction.
\begin{enumerate}
    \item The basic component of our model, built with two layers of STAgg, without significance selection and intimate window moduel (BGNN).
    \item Basic GNN framework with significant selection moduel (BGNN+S). 
    \item Basic GNN framework with intimacy window moduel (BGNN+I)
    \item STGNN with significant selection and intimacy window mechanism (STGNN).
\end{enumerate}
\begin{figure}[ht]
    \centering
    \includegraphics[width=\linewidth]{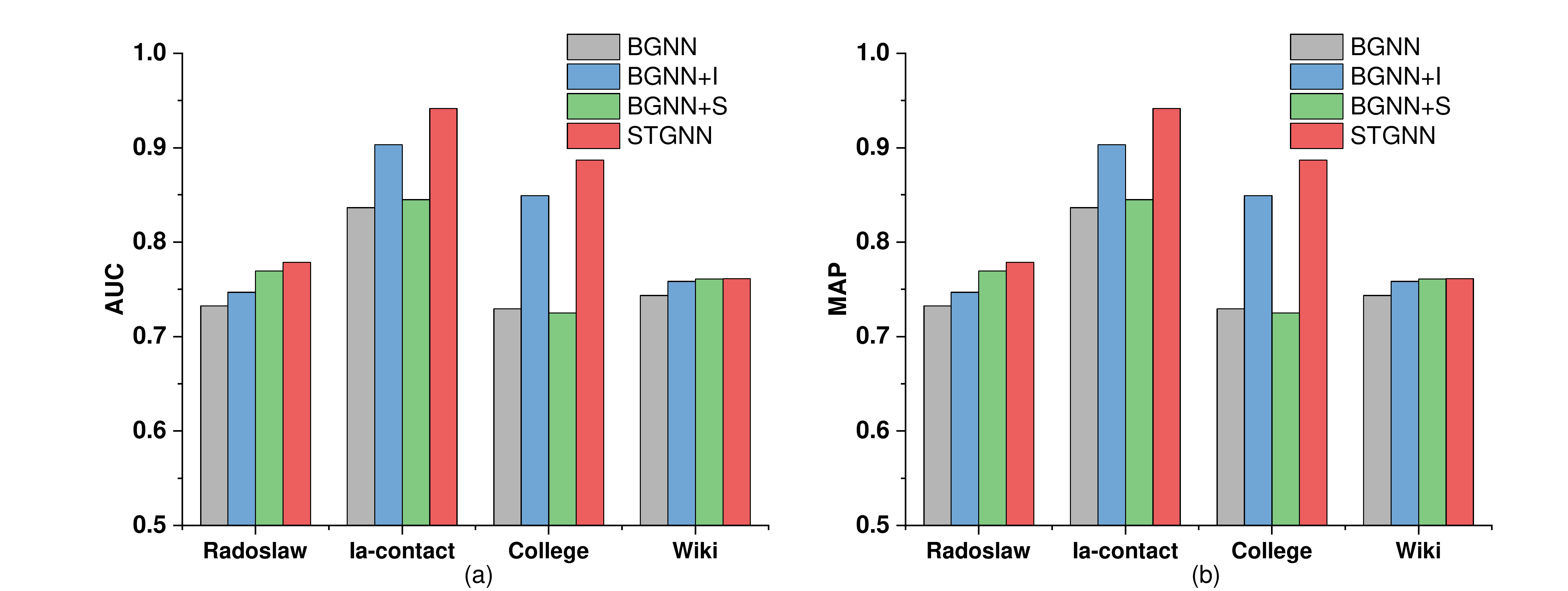}
    \caption{The role of each modules}
    \label{fig: ablation}
\end{figure}
As shown in Fig. \ref{fig: ablation}. As modules continue to be added,  the performance improves. BGNN is the fundamental module of our methodology, aggregating neighbor information based on the significance of the relationship between entities, laying an excellent basic performance of our approach. The intimate window (BGNN+I) brings the most significant performance improvement, validating the effectiveness in capturing and modeling the significance of node pairs. The significant selection (BGNN+S) also brings a performance boost. Indicate this module can improve the information density under a given historical capacity. When integrating both intimate window and significant selection, the whole model STGNN achieves the best performance, showing that more efficient features extract can model more fine-grained relationships between entities.
\subsection{Hyperparameter study}
\begin{figure}[ht]
    \centering
    \includegraphics[width=\linewidth]{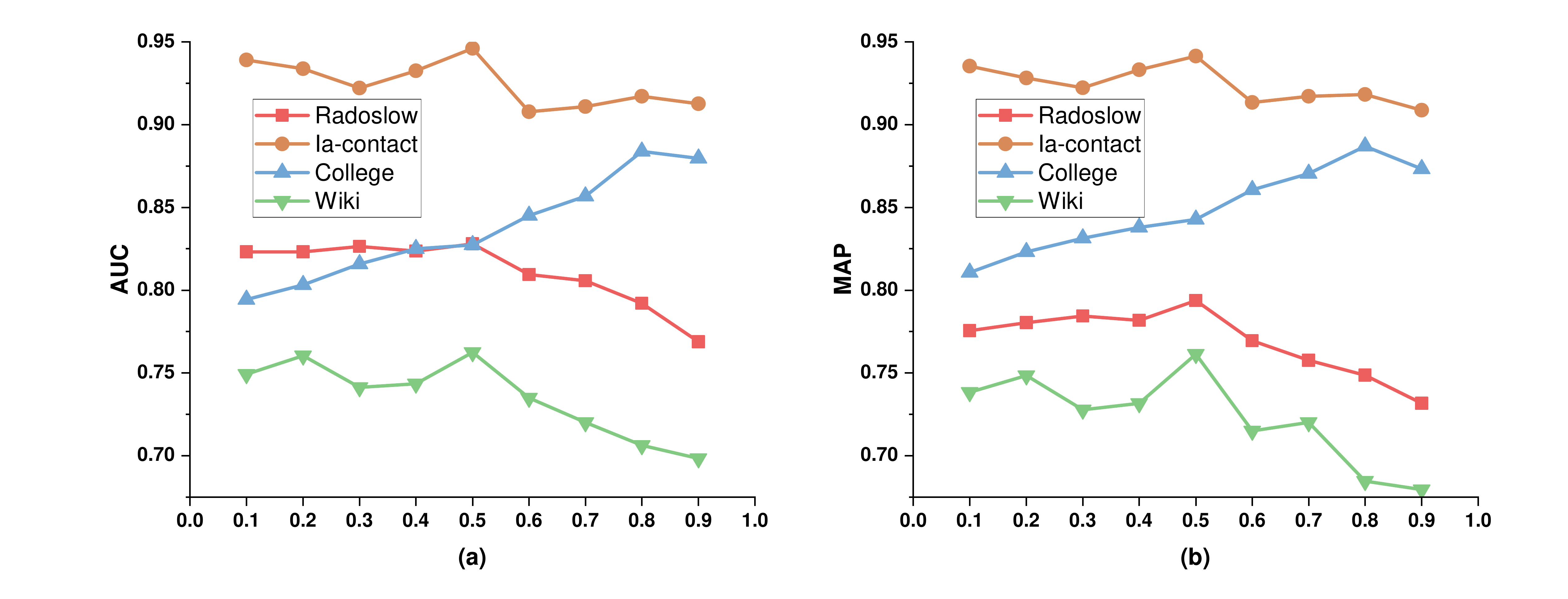}
    \caption{The role of each modules}
    \label{fig: prameters}
\end{figure}
\begin{figure}[h]
    \centering
    \includegraphics[width=\linewidth]{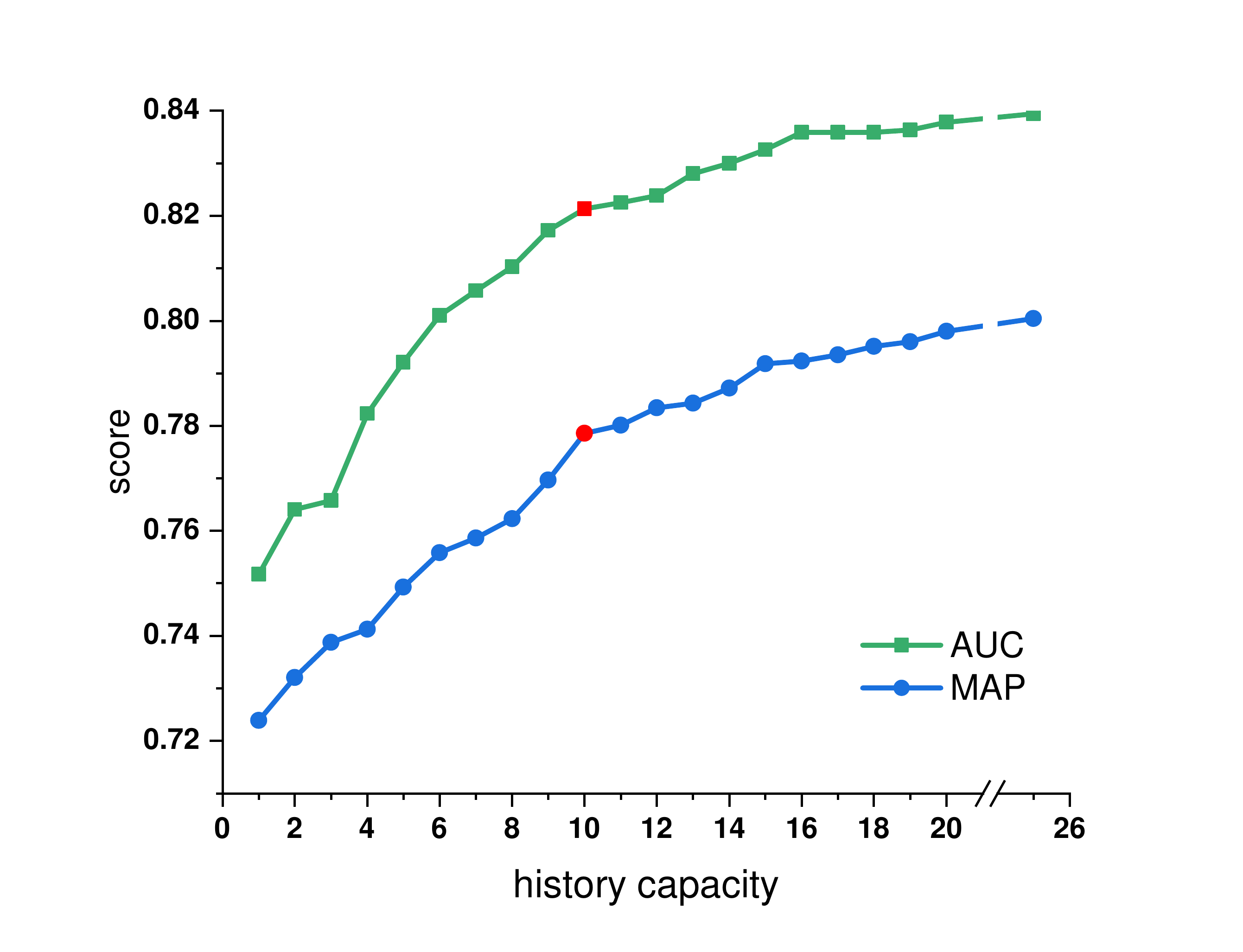}
    \caption{Correlation of performance with history capacity (Radoslaw)}
    \label{fig: history capacity}
\end{figure}
We separately explore the hyperparameter $p$, which controls the size of the intimate window, and the hyperparameter $m$, which controls the history capacity. 
\par As shown in Fig. \ref{fig: prameters}.
Different networks have different sensitivity to the size of the intimate window. Larger intimate windows may lead to performance degradation, as longer observations may generate noise by counting another event cluster to the current moment, while shorter observations may not adequately capture the significance of the moment. According to our subjective experience, setting the size of $p$ between 0.5 and 0.8 will be a universality value.
\par The correlation between performance and historical capacity is shown in Fig. \ref{fig: history capacity} The two red dots are the performance reported by our comparative experiments with the history capacity $m=10$, but when $m$ increases, the performance can be further improved.  We can speculate that the final performance of STGNN on the Radoslaw network will be around AUC=0.84 and MAP=0.80. To our surprise, the high-density neighbor information selected according to significance has shown excellent predictive power even with a small historical capacity. Performance converges quickly as historical capacity increases.  A larger historical capacity will bring specific performance improvements, but its memory overhead increases exponentially. To verify the effectiveness of our significance selection for improving information density, we set a small history capacity for all datasets for comparison. 
\section{conclusion}
In this paper, we propose a novel temporal GNN-based framework STGNN for modeling continuous-time temporal networks. It models the significance of each relation of entities by extracting a more density neighbor historical information. Our experiments show that modeling the level of intimacy between entities is more suitable for the dynamic process of interaction, and the neighbor portraits of nodes are mainly composed of significant neighbors.

\bibliographystyle{unsrt}
\bibliography{refs}
\end{document}